# Current-induced asymmetric magnetoresistance due to energy transfer via quantum spin-flip process

K. –J. Kim[1], T. Moriyama[1], T. Koyama[2], D. Chiba[2], S. –W. Lee[3], S. –J. Lee[4], K. –J. Lee[3,4*], H. –W. Lee[5*], & T. Ono[1*]

[1]Institute for Chemical Research, Kyoto University, Gokasho, Uji, Kyoto, 611-0011, Japan
[2]Department of Applied Physics, Faculty of Engineering, The University of Tokyo, Bunkyo, Tokyo 113-8656, Japan
[3]Department of Materials Science and Engineering, Korea University, Seoul 136-701, Republic of Korea
[4]KU-KIST Graduate School of Converging Science and Technology, Korea University, Seoul 136-713, Republic of Korea
[5]PCTP and Department of Physics, Pohang University of Science and Technology, Pohang 790-784, Korea

*Correspondence to: kj_lee@korea.ac.kr, hwl@postech.ac.kr, ono@scl.kyoto-u.ac.jp

Current-induced magnetization excitation is a core phenomenon for next-generation magnetic nanodevices[1], and has been attributed to the spin-transfer torque (STT) that originates from the transfer of the spin angular momentum between a conduction electron and a local magnetic moment through the exchange coupling[2–11]. However, the same coupling can transfer not only spin but also energy[12–20], though the latter transfer mechanism has been largely ignored. Here we report on experimental evidence concerning the energy transfer in ferromagnet/heavy metal bilayers. The magnetoresistance (MR) is found to depend significantly on the current direction down to low in-plane currents, for which STT cannot play any significant role. Instead we find that the observed MR is consistent with the energy transfer mechanism through the quantum spin-flip process, which predicts short wavelength, current-direction-dependent magnon excitations in the THz frequency range. Our results unveil another aspect of current-induced magnetic excitation, and open a channel for the dc-current-induced generation of THz magnons.

When a conduction electron is injected into a ferromagnet (FM), its spin $\sigma$ interacts with the spin $S$ of the local magnetic moment in the FM through the exchange coupling $-J_{ex}\sigma \cdot S$. Here, $J_{ex}$ is the Heisenberg exchange coupling constant. When $S$ is regarded as a classical spin, the coupling does not induce any spin dynamics if $\sigma$ is perfectly parallel or antiparallel to $S$. On the other hand, when $\sigma$ has a transverse component with respect to $S$, $\sigma$ experiences dynamic evolution; owing to the conservation of spin angular momentum, this induces a torque on $S$ [Fig. 1a]. This constitutes the spin-transfer torque (STT) mechanism[2–3], which has successfully explained various experimental results concerning current-induced magnetization dynamics[4–11]. Nevertheless, it has been suggested that the orthodox theory of STT may be incomplete[12] because it neglects energy transfer and resulting generation of finite-wavelength magnons[13]. However experimental observation of such magnons remains elusive. Early claims[14–15] of such observation were disputed[6].

The mechanism of the energy transfer becomes evident when both $\sigma$ and $S$ are treated quantum mechanically. When the local magnetic moment lies in the $z$ direction, the exchange coupling becomes $-J_{ex}\sigma \cdot S = -J_{ex}(\sigma_z S_z + \sigma_+ S_-/2 + \sigma_- S_+/2)$, where $\sigma_\pm = \sigma_x \pm i\sigma_y$ and $S_\pm = S_x \pm iS_y$ are creation/annihilation operators for the spin of the conduction electron and the spin of the local magnetic moment, respectively. The first term of the exchange coupling, $-J_{ex}\sigma_z S_z$, is the Zeeman energy, and is the only term that is taken into account when orthodox theories of STT deal with conduction electron dynamics. The next two terms, $-J_{ex}(\sigma_+ S_- + \sigma_- S_+)/2$, arise from the quantum uncertainties of both $\sigma$ and $S$ (which have been ignored in orthodox STT theories), and do not generate any dynamics when $\sigma$ and $S$ are parallel. However, when they are anti-parallel, the two terms flip the spin of the conduction electron and simultaneously generate a magnon [Fig. 1b]. This quantum spin-flip process occurs when the STT mechanism does not predict that any torque arises. It thus goes beyond the predictions of the STT mechanism.

To clarify its connection with the energy transfer, we note that the quantum spin-flip process involves a conduction electron jump from a minority to a majority energy band [Figs. 1c and 1d], which inevitably involves momentum transfer from the conduction electron to the magnon because the Fermi wavevectors $k_{F,maj}$ and $k_{F,min}$ of the majority and minority bands differ from each other. Thus the magnon carries a momentum of at least $\hbar(k_{F,max} - k_{F,min})$. For conventional FMs such as Co and Fe[16,17], the relevant $k_{F,max} - k_{F,min}$ is on the order of $0.1 \times \pi/a \sim 10^9$ m$^{-1}$ (where $a$ is the atomic spacing), which provides a lower bound of the magnon wavevector. According to high-frequency magnon dispersion measurements[18,19], the lower bound of the magnon wavevector corresponds to a magnon energy of the order of 5 meV, which amounts to 1.2 THz. Thus the quantum spin-flip process involves energy transfer, and the resulting magnon can have a frequency of ~1 THz, which is about three orders of magnitude larger than the frequency of typical current-induced magnetization dynamics (i.e. frequencies on the order of GHz) probed in earlier experiments[4–11]. This may be the reason why STT theories, despite their incompleteness, can explain the low-frequency (or long-wavelength) experiments so successfully.

Although the quantum spin-flip process has been neglected in STT-related studies, its existence has been evidenced in several experiments on magnetic excitation by hot electrons, such as spin-polarized high-resolution electron energy loss spectroscopy (SPEELS) [18-19], two-photon photoemission measurements[19], and inelastic scanning tunnelling spectroscopy[13]. An important outstanding question is whether the quantum spin-flip process is relevant for dc-current injection, which involves much lower energy scale than SPEELS and is typical experimental configuration for STT-related studies. If this is the case, then it allows for a considerable extension of our understanding of current-induced magnetic excitation; more importantly, it lays the foundation for practical THz spintronics devices operated by a dc current. As far as energy scale is concerned, this possibility is plausible since the magnon energy for frequency $v = 1$ THz is only $hv = 4$ meV. Hence not only energetic (~5 eV) electron beams such as those in SPEELS, but also conduction electrons with much lower excess energy above the Fermi energy, may also excite magnons in the THz range.

To check if dc current can excite magnons in the THz range, we choose electron-magnon scattering as a detection tool, which is probed by the dc electrical resistance measurement[21,22]. The reason for this choice is threefold. First, it is extremely challenging to electrically detect THz signals in the frequency domain when a dc current is injected into a nanodevice. Second, the contribution of the electron-magnon scattering to the electrical resistance is expected to be dominated by magnons excited by the quantum spin-flip process, because they have much larger momenta (or shorter wavelength) than magnons excited by the STT mechanism, and magnons with larger momenta cause stronger momentum scattering of electrons. Third, the dc resistance measurement can probe the directional dependence of the magnon population in the THz range. According to the quantum spin-flip process, the magnon population is maximally enhanced (largest resistance) when $\sigma$ is antiparallel to $S$. On the other hand, when $\sigma$ is parallel to $S$, the magnon population is maximally suppressed (smallest resistance) since the quantum spin-flip process then annihilates magnons in the THz range [Fig. 1b]. Therefore, the directional dependence of the dc electrical resistance can unambiguously provide evidence of the quantum spin-flip mechanism.

Figure 2a shows a schematic diagram of a wire structure composed of the $Ni_{80}Fe_{20}$ (=Py, 5 nm) / Pt (5 nm) bilayer that we used for the experimental test. The width and length of the wire are 100 nm and 20 μm, respectively, and the magnetic easy axis of the bilayer lies along the $x$ axis. Electric current is injected along the $+x$ direction, which we define as the positive current. Because of the spin Hall effect in Pt[23], an in-plane charge current generates an out-of-plane spin current, which injects conduction electron spins into the ferromagnetic Py layer. The red and blue arrows in Fig. 2a represent the spin of conduction electrons, whose trajectories are deflected by the spin Hall effect. Pt has a positive spin Hall angle, which we verify for our bilayer by a spin-torque ferromagnetic resonance (ST-FMR) measurement[24]. Hence a positive in-plane current injects conduction electrons with their spin along the –y direction into the Py. To test the directional dependence of the electrical resistance, we sweep the magnetic field $B$ along the $y$ direction. For sufficiently large positive (negative) values of $B$, $S$ in the Py lies along the –y (+y) direction, and $S$ is parallel (antiparallel) to $\sigma$ for a positive current density $J$.

Figure 2c shows the variation of the longitudinal resistance $R$ as a function of $B$ for various positive values of $J$. The vertical axis represents the normalized resistance $\Delta R/R_{max}$, where $R_{max}$ denotes the maximum value of $R$ during the $B$ scan and $\Delta R(B)=R(B)-R_{max}$. For a small current (black line), $\Delta R/R_{max}$ as a function of $B$ exhibits nearly symmetric $B$ dependence with respect to $B = 0$. This symmetric feature agrees with the expected behaviour for anisotropic magnetoresistance (AMR). However, for a larger current, we find that the symmetry is broken substantially. In the large |B| part of the graph (i.e. >50 mT), $\Delta R/R_{max}$ increases rapidly with $J$ only for negative values of $B$ ($S$ and $\sigma$ are antiparallel), whereas such rapid increase does not occur for positive values of $B$ ($S$ and $\sigma$ are parallel). The sign dependence of $\Delta R/R_{max}$ on $B$ qualitatively agrees with the mechanism of the quantum spin-flip process. For negative values of $J$, on the other hand, the quantum process predicts an enhanced $\Delta R/R_{max}$ for positive values of $B$, because the spin direction of the injected conduction electrons is reversed [Fig. 2b]. The experimental data in Fig. 2d show that this is indeed the case. The directional dependence is also confirmed by other combinations of materials with different spin Hall angles [See Supplementary Information].

Figure 2e summarizes the $J$ dependence of $\Delta R/R_{max}$. Part of the $J$ dependence is due to Joule heating in combination with the temperature dependence of the resistance. To remove this undesired effect, which is an even function of $J$, we focus on the odd component of $\Delta R/R_{max}$ by introducing the asymmetry parameter $r_{asym}(B, J)\equiv[\Delta R(B, -J)-\Delta R(B, J)]/R_{max}$ (we refer to this as the asymmetric MR), which equals the unidirectional MR $[\Delta R(-B, J)-\Delta R(B, J)]/R_{max}$ reported in a recent experiment[25], because $\Delta R(-B,J) = \Delta R(B,-J)$. In Fig. 2f, we plot $r_{asym}(B = +300$ mT, $J)/R_{AMR}$ as a function of $J$, where $R_{AMR}$ is the AMR amplitude. Figure 2f shows that the asymmetric MR is linear in $J$ up to $J_{th}$ (=1.5×10$^{12}$ A/m$^2$) – which we later identify with the STT threshold – and increases more rapidly above $J_{th}$. We note that other types of MR possibly emerging in FM/heavy metal structures – such as spin Hall MR[26], hybrid MR[27], and anisotropic interface MR[28] – are qualitatively different from the asymmetric MR because their effects are even in $J$, whereas $r_{asym}(B, J)$ is odd in $J$.

In what follows, we show that STT (or the damping-like component of the spin-orbit torque) cannot capture the current-induced asymmetric MR. We note that STT generates not only the FMR mode (i.e. magnons with zero wavevectors) but also spin waves (i.e. magnons with a small wavevector) because of sample shape-dependent inhomogeneous magnetic fields coupled with thermal fluctuations[5,29]. For the configuration in Fig. 2a (Fig. 2b), STT enhances (reduces) the effective damping, which can make the number of magnons dependent on the current polarity, just as the quantum spin-flip process does. To test this possibility, we performed micromagnetic simulations of the Landau-Lifshitz-Gilbert (LLG) equation for the device structure in Figs. 2a and 2b, with STT included at a temperature of 300 K (see Methods). The simulations however neglect the quantum spin-flip process. Figure 2g shows the computed asymmetric magnon number $\Delta N_m(J)$ ($\equiv N_m(+J) - N_m(-J)$), where $N_m$ ($= <M_S - M_y>/M_S$) is proportional to the difference between the saturation magnetization $M_S$ and the $y$ component of the magnetization ($M_y$). Above $J_{th}$, where the net effective damping becomes negative for one current polarity, a sizable value of $\Delta N_m(J)$ is obtained. However, for $|J| < J_{th}$, $\Delta N_m(J)$ is very negligible; this is in contrast to the experimental data of Fig. 2f, which show a significantly asymmetric MR even below $J_{th}$. Hence we conclude that the linear-in-$J$ dependence of $r_{asym}(B, J)$ for $|J| < J_{th}$ cannot be explained by the STT mechanism. Interestingly, the LLG simulation results in Fig. 2g agree well with Fig. 2h, which is obtained by subtracting the linear contribution (green line) from the experimental data shown in Fig. 2f. Hence we attribute the nonlinear $J$ dependence of $r_{asym}(B, J)$ above $J_{th}$ to the STT mechanism.

Time-resolved measurements of the asymmetric MR provide another means to distinguish between the energy transfer and spin transfer effects. Figure 3a shows time-resolved variation of the asymmetric MR where the current is turned on at $t = 0$ [for details of the experiment, see Methods]. From the fit based on $1 - \exp(-t/\tau)$ for $t > 0$, we obtain the characteristic time $\tau$ for each value of $J$. The experimental results [Fig. 3c; the first three points are for below $J_{th}$ (see Fig. 3b)] show a monotonic decrease of $\tau$ with increasing $J$. On the other hand, the micromagnetic simulation of spin transfer effects results in a rapid increase of $\tau$ up to $J \sim J_{th}$, in clear contrast to experimental results [Fig. 3d].

To obtain insight into this phenomenon, it is useful to discuss the physical meaning of $\tau$ when $\sigma$ and $S$ are antiparallel. In the simulation of the STT effects, $\tau$ is the time scale for the transition from an initial magnetic state to an excited state. Because these two states have different energies, $\tau$ is determined by the inverse of the energy dissipation rate – i.e. 1/|effective damping|. Considering that the net effective damping decreases with $J$ until it becomes zero at $J = J_{th}$, $\tau$ should exhibit a peak at $J = J_{th}$ according to the STT mechanism, as supported by the numerical calculation results [Fig. 3d]. This $J$ dependence of $\tau$ is very robust. Hence the different behaviour of $\tau$ between Figs. 3c (experiment) and 3d (STT modelling) demonstrates that the origin of the asymmetric MR markedly differs from the predictions of the STT mechanism.

It was recently suggested that an asymmetric MR may arise if the spin Hall effect modifies the scattering potential at the interface in a current-polarity-dependent manner, which in turn modifies the spin accumulation at the interface[25,30]. Although such spin accumulation may indeed occur, it cannot be the main reason for the asymmetric MR. If the THz magnons were absent and only GHz magnetization dynamics were induced by a dc current, the spin transport (on the order of ps) is much faster than the magnetization dynamics (on the order of ns). Then the characteristic time scale over which the spin accumulation develops is governed by the magnetization dynamics. Therefore, in the absence of any mechanism that modifies the magnetization dynamics, the $J$-dependence of $\tau$ arising from the spin accumulation should be bound to the modelling results shown in Fig. 3d. Hence, in order to explain the experimentally obtained $\tau$ [Fig. 3c], the nature of the magnetization dynamics itself should be altered.

In this regard, the quantum spin-flip mechanism naturally explains the experimental result of $\tau$ as follows. We demonstrate this for the current polarity, for which dc current generates magnons. Because the generation of magnons in the THz range requires energy transfer through the quantum spin-flip mechanism, the magnon generation rate by a dc electric current is proportional to the energy dissipation rate $JV \propto J^2$ of electric current, where $V$ is the voltage applied to generate current. On the other hand, the magnon relaxation rate is of the form $N/\tau$, where $N$ is the THz range magnon number. Therefore, the overall rate is given by $dN/dt = AJ^2 - BN/\tau$, where $A$ and $B$ are positive constants. Then the steady-state value $N_{\text{steady}}$ of $N$ satisfies $BN_{\text{steady}} = \tau AJ^2$. Assuming that $r_{\text{asym}}(B, J)$ measured at the steady state is linear in $N_{\text{steady}}$, and considering that $r_{\text{asym}}(B, J)$ is linear in $J$ for $J < J_{\text{th}}$, one finds that $\tau$ should be proportional to $1/J$, which is indeed the case for the first three points in Fig. 3c, corresponding to the data obtained for $J < J_{\text{th}}$. As a passing remark, we mention that the time-resolved measurement also allows one to exclude the anomalous Nernst effect from a possible mechanism of the asymmetric MR (Supplementary Information).

The angular dependence of the asymmetric MR also supports the quantum spin-flip process, which predicts $-\cos\theta$ dependence, where $\theta$ is the angle between the conduction electron spin $\sigma$ and the magnetization spin $S$, because the process generates (annihilates) magnons maximally when $\sigma$ and $S$ are antiparallel (parallel). As shown in Fig. 4a, the experimental results obtained by rotating the field $B$ in the $xy$, $yz$, and $zx$ planes are consistent with this prediction. We also check the temperature dependence of the asymmetric MR. Interestingly, the asymmetric MR increases with increasing temperature $T$ [Fig. 4c], which is in stark contrast to the AMR [Fig. 4b] and other MRs, such as giant MR, which become weaker at elevated temperatures. Considering that the magnon relaxation becomes faster at larger $T$, the result in Fig. 4c implies that the magnon excitation mechanism becomes more effective at larger $T$. This is consistent with the quantum spin-flip process, for which the thermal energy facilitates the excitation of magnons.

Finally we experimentally estimate the magnon frequency arising from the quantum spin-flip process. Figure 5 shows the asymmetric MR as a function of $B$ up to 9 T, which is the maximum field that we can apply in our setup. Measurement temperature was set to 10 K to avoid the thermal magnon effect. The asymmetric MR does not change in these field ranges, meaning that the energy scale of the quantum process is larger than 1 meV (computed from $g\mu_B B$ with $g = 2$), which in turn implies the magnon frequency to be at least 0.25 THz. Therefore, this result shows that the dc-current-induced quantum spin-flip process generates magnons in much higher frequency ranges than the STT does.

Our study reveals the energy transfer effect of the dc spin current on the magnetization dynamics. This effect has been ignored for more than a dozen years because most STT experiments have focused on magnetization dynamics in the GHz range, which is far below the frequency range relevant to the energy transfer effect. Because the quantum spin-flip process allows for the generation of magnons in the THz range by a simple dc current without any high-energy electron beam source or sophisticated excitation techniques, it may pave the way towards realizing a convenient THz magnon source and ultrafast electro-magnetic devices. In addition, this dc-current-induced quantum spin-flip process becomes stronger with increasing temperature and is clearly visible at room temperature, even though it arises from quantum uncertainty. Therefore, this work opens an avenue towards studying quantum uncertainty at room temperature. It is also interesting to investigate possible connections and interplay between this result and laser-induced ultrafast magnetization dynamics [31–32], which also falls into THz frequency range.

## Methods

**Experiment.** Several kinds of ferromagnet (FM)/non-magnet (NM) bilayer films were prepared using rf magnetron sputtering. The $Ni_{80}Fe_{20}$ (Py), Co and Fe were used for FM and Pt, Ta and Cu were used for NM. The films were patterned into the wires by electron beam lithography and Ar ion milling. The width of wire ranged from 100 nm to 10 μm and the length of wire ranged from 1 μm to 100 μm. The thickness of each layer was modulated in the range of 1 nm ~ 10 nm. We confirmed that all the samples show the asymmetric MR that is predicted by quantum spin-flip process. The transport measurement shown in Fig. 2 was done by using a dc source (Yokogawa 7651, max: 30 V, 100 mA) and nanovoltmeter (Keithley 2182) at room temperature ($T$ = 295 K). A time-resolved measurement shown in Fig. 3 was performed at a high frequency probe system by connecting pulse generator (Picosecond 10,300B, +50/−45 V with rising time < 0.3 ns) and real-time oscilloscope (DPO 7354) serially with sample. Thus, the voltage pulse $V_P$ is injected from the pulse generator into the sample and the outgoing voltage pulse $V_O$ is recorded by the oscilloscope. To obtain the asymmetric MR, i.e., the difference in resistance for ± $B$, we subtracted the recorded pulse for +$B$ from that for −$B$. To get an enough signal-to-noise ratio, the subtracted pulse was averaged by 1,000 times repeated measurements (for first three data (black, red and green data in Fig. 3a), we repeated 3,000 times due to their weak signal). The resistance of sample was calculated by the circuit equation $R = R_O(V_P/V_O - 1) - R_F$, where $R_O$ and $R_F$ are the load resistance of the oscilloscope and the pulse generator, respectively[33]. The current strength was calculated from the recorded voltage and load resistance of oscilloscope (50 Ω), which was then converted to the current density by assuming equal distribution of current in bilayer. High field measurement and low temperature measurement, shown in Fig. 4 and Fig. 5, were performed using physical property measurement system (Quantum Design, 4.3K–350K, up to 9T).

**Micromagnetic simulation.** Micromagnetic simulation was performed by numerically solving the Landau-Lifshitz-Gilbert equation including the damping-like spin-orbit torque, given as $\partial \hat{\mathbf{m}}/\partial t = -\gamma \mu_0 \hat{\mathbf{m}} \times \mathbf{H}_{eff} + \alpha \hat{\mathbf{m}} \times \partial \hat{\mathbf{m}}/\partial t + \gamma(\hbar/2e)(J_e/M_S d)\theta_{SH}\hat{\mathbf{m}} \times (\hat{\mathbf{m}} \times \hat{\mathbf{y}})$. Here, $\hat{\mathbf{m}}$ is the unit vector along the magnetization, $\gamma$ is the gyromagnetic ratio, $\mathbf{H}_{eff}$ is the effective magnetic field including the exchange, magnetostatic, external, and thermal fluctuation fields, $\alpha$ is the damping constant, $e$ is the electron charge, $J_e$ is the electric current density, $M_S$ is the saturation magnetization, $d$ is the thickness of FM, and $\theta_{SH}$ is the effective spin Hall angle. The following parameters were used: The width and thickness of wire are 100 nm and 5 nm, respectively, the saturation magnetization $M_S$ is 800 kA/m, the exchange stiffness constant is $1.3 \times 10^{-11}$ J/m, the spin Hall angle $\theta_{SH}$ is 0.07, the intrinsic damping constant is 0.05, the temperature is 300 K, and the unit cell size is $4 \times 4 \times 5$ nm$^3$. For a finite temperature with the stochastic calculation, the Gaussian-distributed random fluctuation fields (mean = 0, standard deviation = $\sqrt{2\alpha k_B T/(\gamma M_S V \Delta t(1+\alpha^2))}$, where $\Delta t$ is the integration time step, $V$ is the volume of unit cell) were considered in the effective field of the Landau-Lifshitz-Gilbert equation.

The number of magnons shown in Fig. 2g is the value of difference between $(1-|<m_y>|)$ at $B$ = +300 mT and $B$ = −300 mT. The averaged $y$-component of magnetization unit vector $<m_y>$ was obtained by averaging $m_y$ over the modelled wire for 5 ns. The characteristic time $\tau$ shown in Fig. 3d was obtained by fitting the temporal evolution of the value of $(1-|<m_y>|)_{B = +300 \text{ mT}} - (1-|<m_y>|)_{B = +300 \text{ mT}}$ with the equation of $1-\exp(-t/\tau)$.

## Acknowledgements

We thank M. D. Stiles and Y. Tserkovnyak for fruitful discussion. K.-J.K., T.M. and T.O. acknowledge support from JSPS KAKENHI Grant Numbers 15H05702, 26870300, 26870304, 26103002, Collaborative Research Program of the Institute for Chemical Research, Kyoto University, and R & D Project for ICT Key Technology of MEXT from the Japan Society for the Promotion of Science (JSPS). T.K. and D.C. were supported by JSPS KAKENHI Grant Numbers 25220604. S.-W.L., S.-J.L. and K.-J.L. acknowledge support from the National Research Foundation of Korea (NRF) (2011-028163, NRF-2013R1A2A2A01013188). H.-W.L. acknowledges support from NRF (2013R1A2A2A05006237).

## Author contributions

K.-J.K., K.-J.L., H.-W.L. and T.O. conceived the idea and planned the study. T.K. and D.C. provided films. K.-J.K. fabricated the devices and performed the experiment. T.M. contributed ST-FMR measurement. S.-W.L., S.-J.L. and K.-J.L. performed micromagnetic simulation. K.-J.K., K.-J.L., H.-W.L. and T.O analysed the data and wrote the manuscript. All authors discussed the results.

## Additional information

Supplementary information is available in the online version of the paper. Reprints and permissions information is available online at www.nature.com/reprints. Correspondence and requests for materials should be addressed to K.-J.L., H.-W.L., and T.O.

# Figure Legends

**Figure 1| Schematic illustration of spin transfer torque (STT) and quantum spin-flip process. (a)** Spin-transfer torque (STT) process. **(b)** Quantum spin-flip process. Black arrows represent the spin $S$ of the local magnetic moment. Red and blue arrows represent the spin $\sigma$ of the conduction electron. **(c,d)** Schematic illustration of energy bands for magnon **(c)** creation and **(d)** annihilation via the quantum spin-flip process. $E_F$ denotes the Fermi energy of the ferromagnet (FM).

**Figure 2| Current-induced asymmetric magnetoresistance (MR) of Py/Pt bilayer. (a,b)** Schematics of bilayer structure with the definition of external magnetic field ($B$), current ($J$), and spin Hall angle. Red and blue arrows in Pt exhibit the spin $\sigma$ of conduction electrons, which are deflected by the spin Hall effect. Black arrows in Py indicate the spin $S$ of the local magnetic moment. **(c,d)** MR results as a function of magnetic field $B$. Different colours correspond to the different currents whose densities are denoted in the legend. Here, the current density was calculated by assuming a uniform distribution of current in the bilayer. **(e)** The variation of MR as a function of current density $J$ for a positive (red) and negative (blue) magnetic field ($|B|$ = 300 mT). **(f)** Linear component of MR variation, which is extracted from Fig. 2e after subtracting the even component. The solid line is the linear fitting below the threshold. **(g)** A micromagnetic simulation result of time-averaged magnetization fluctuation as a function of $J$. $J_{th}$ is the threshold current density. Material parameters are denoted in the Method. **(h)** A rescaled graph of Fig. 2f after subtracting the linear background.

**Figure 3| Time-resolved measurement of asymmetric magnetoresistance (MR). (a)** Temporal evolution of asymmetric MR. Different colours correspond to the different electric currents whose densities are 0.9 (black), 1.1 (red), 1.4 (green), 1.7 (blue), 1.9 (magenta), 2.1 (purple), 2.3 (wine), 2.5 (olive), and 2.7 (orange) $\times 10^{12}$ A/m². Solid lines are the exponential fit based on $1-\exp(-t/\tau)$. $B$ = 300 mT is applied in the $y$ direction. **(b)** Saturation values of asymmetric MR as a function of current density $J$. The blue solid line denotes the linear background of the asymmetric MR, which is the same as that of Fig. 2f. **(c)**

Characteristic time $\tau$ as a function of $J$. The red solid line represents the $1/J$ fit curve. The error bars represent the fitting error (the size of error bars are smaller than that of symbols). **(d)** $\tau$ as a function of $J/J_{th}$ obtained by micromagnetic simulations.

**Figure 4| Angular dependence and temperature dependence of asymmetric magnetoresistance (MR). (a)** Angular dependence of asymmetric MR. The angle is indicated in the figure inset. The solid lines in the middle and bottom panels are the fitting curves based on $-\cos\theta$. $B = 3$ T and $J = 6 \times 10^{11}$ A/m² are applied. **(b)** Temperature dependence of anisotropic magnetoresistance (AMR). **(c)** Temperature dependence of asymmetric MR. In **(b)** and **(c)**, the data were obtained from Py (10 nm) / Pt (5 nm) for $B = 300$ mT and $J = 9 \times 10^{11}$ A/m².

**Figure 5| External field dependence of asymmetric magnetoresistance (MR).** Asymmetric MR as a function of $B$, which was obtained from Py (3 nm) / Pt (5 nm) at 10 K. Different colours correspond to the different currents whose densities are denoted in the legend.

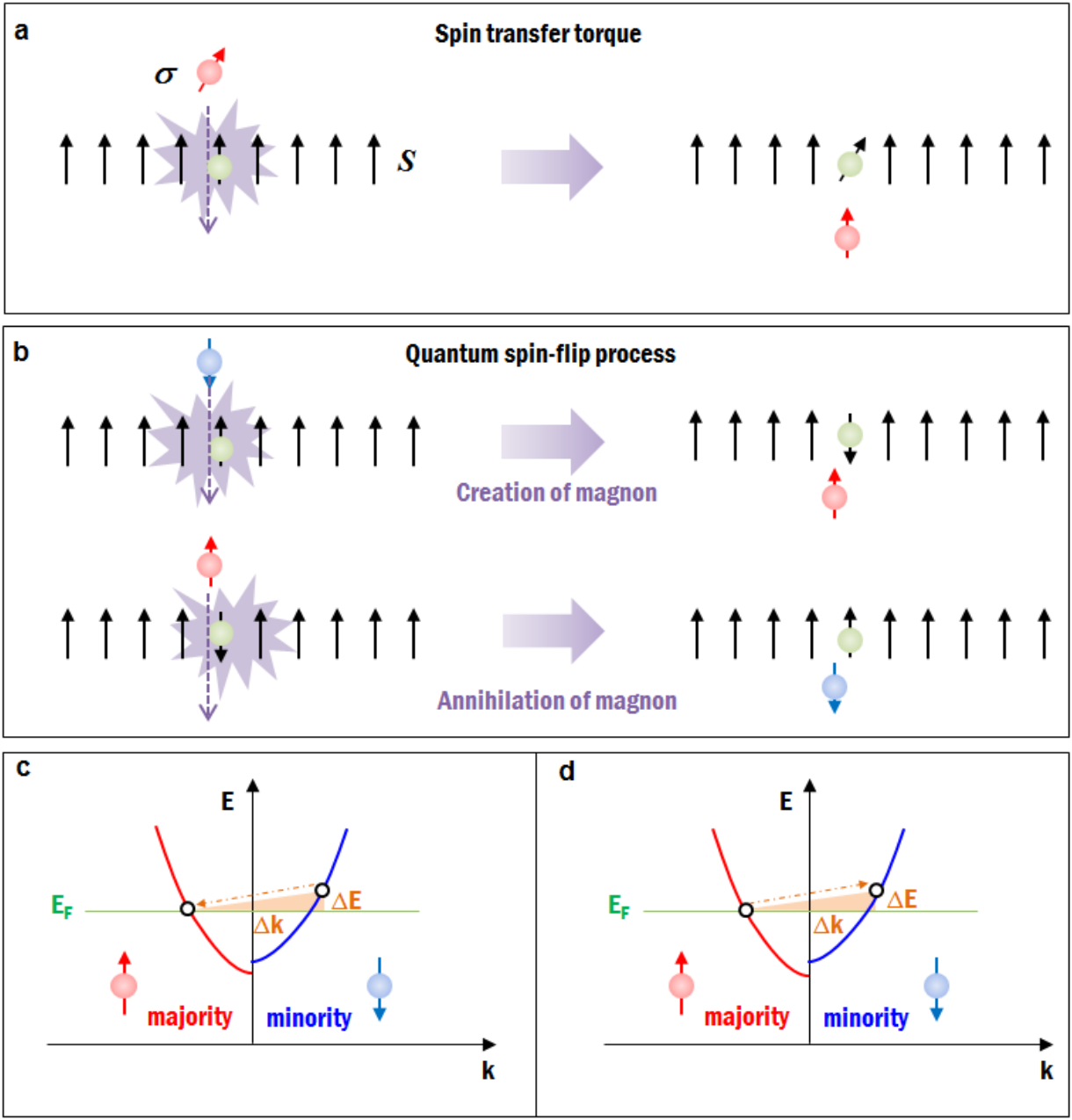

**Figure 1**

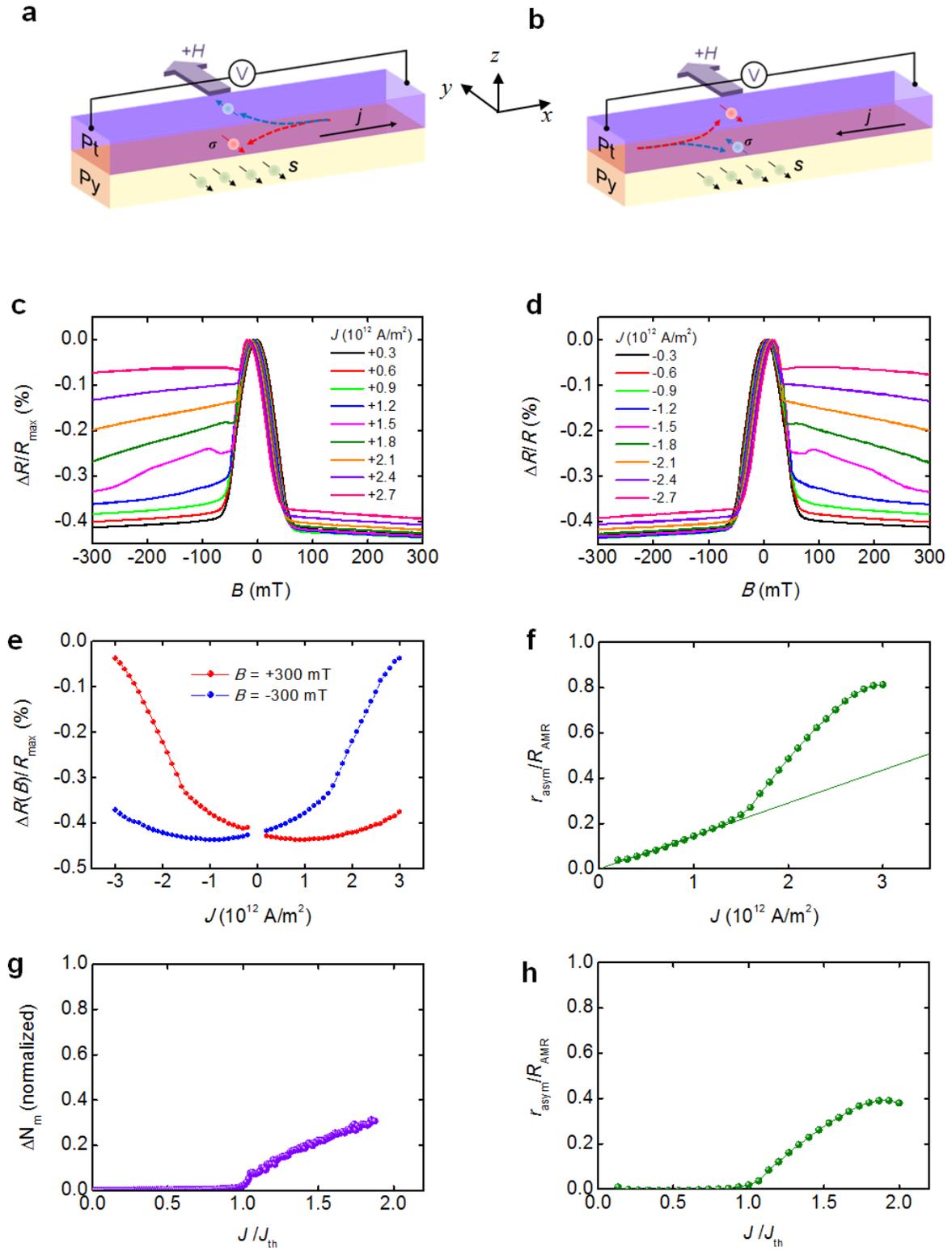

**Figure 2**

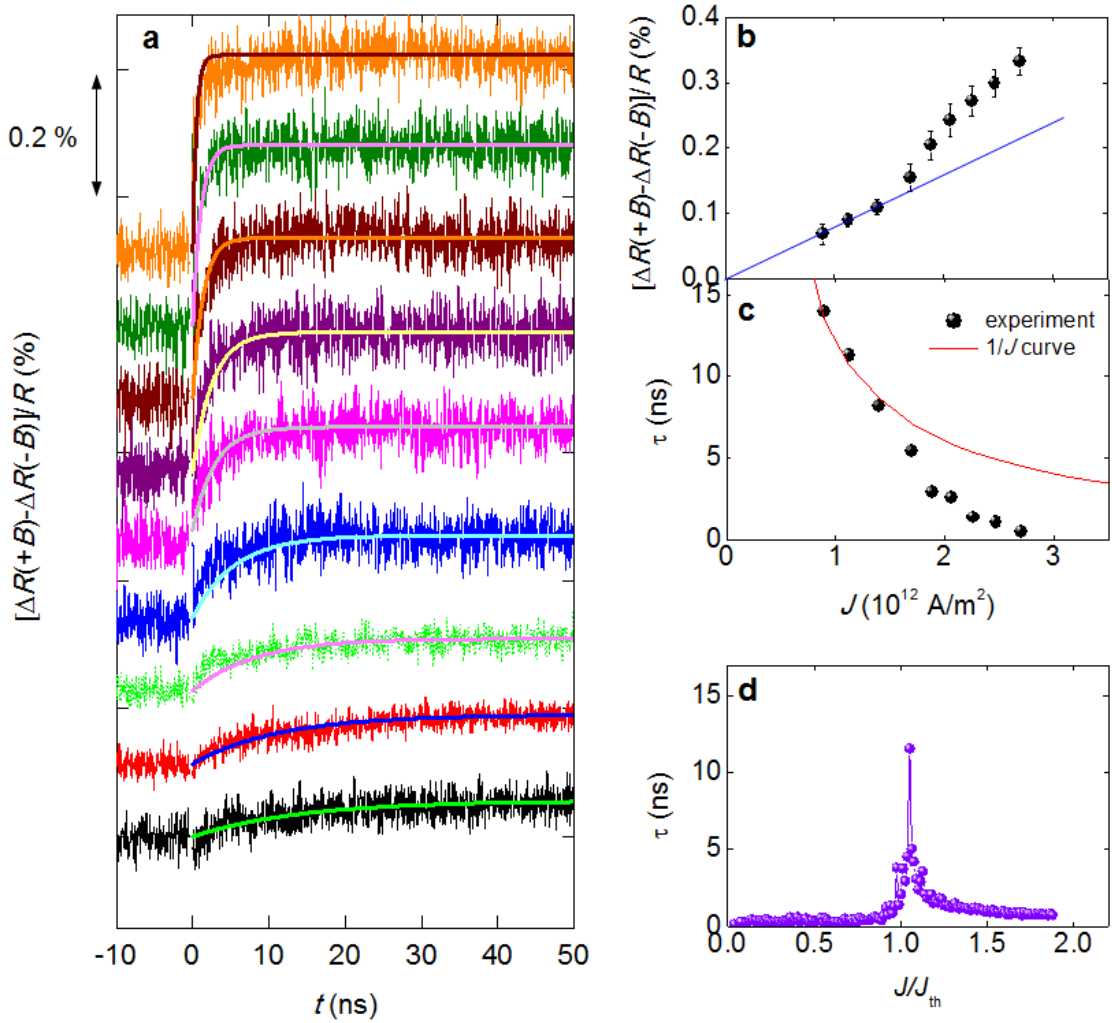

**Figure 3**

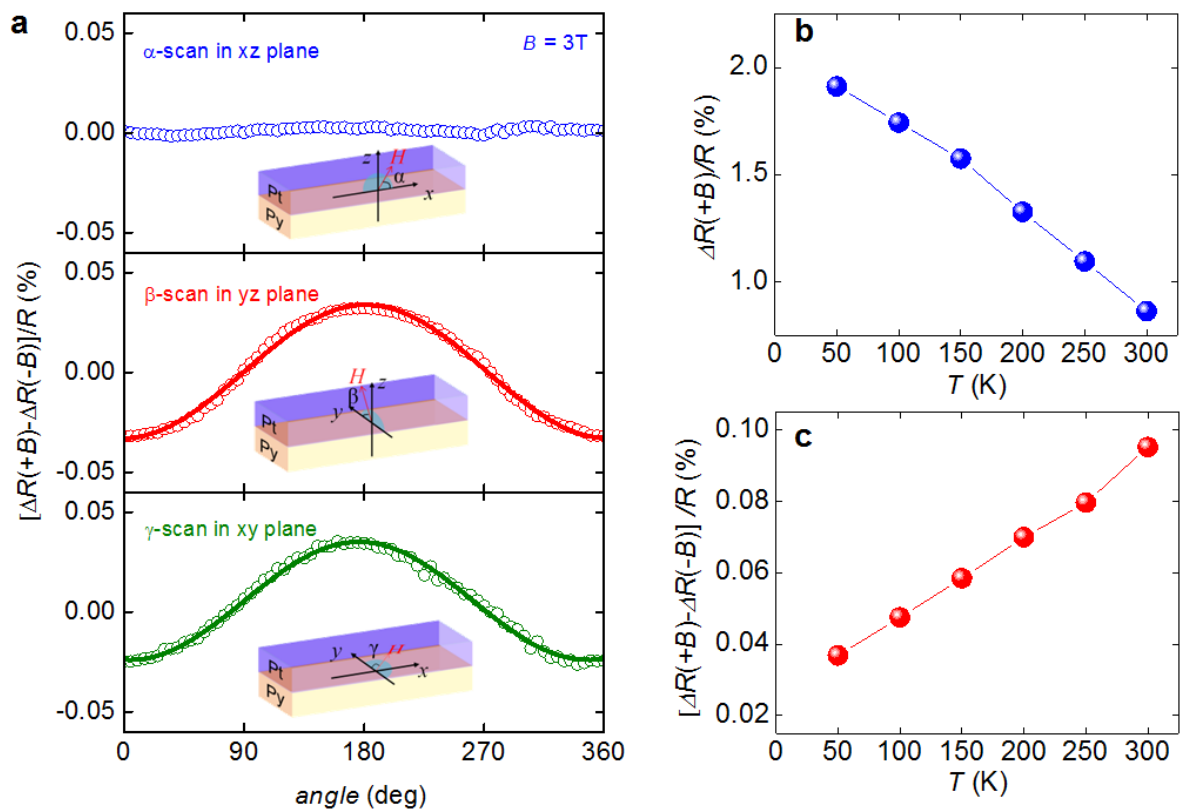

Figure 4

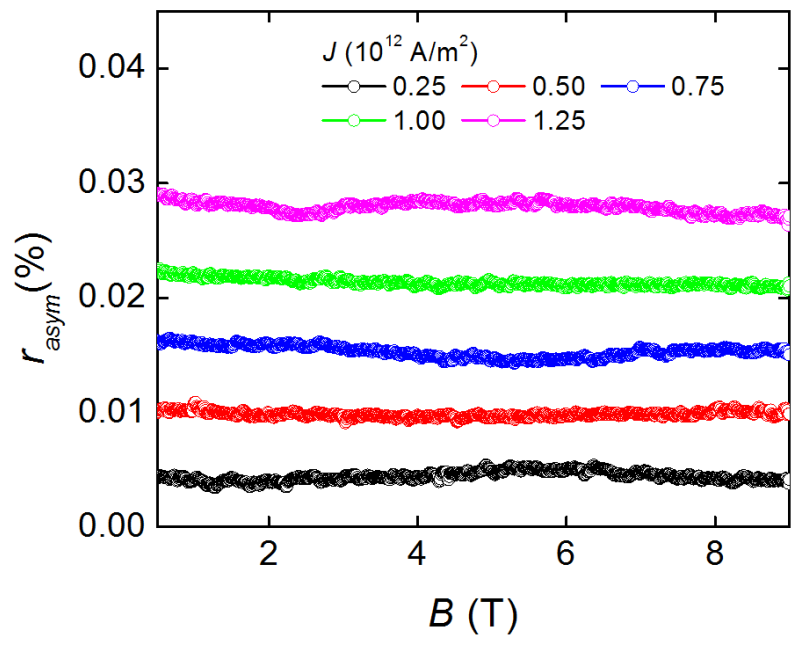

**Figure 5**